\documentclass{article}
\usepackage[preprint]{spconf}
\copyrightnotice{\copyright\ IEEE 2022}
\toappear{To appear in {\it Proc.\ ICASSP 2022, May 22-27, 2022, Singapore}}

\usepackage{amsmath,graphicx}
\usepackage{amssymb}
\usepackage{amsfonts}
\usepackage{mathtools}
\usepackage{multirow}
\usepackage{bm}
\usepackage{nicefrac}
\usepackage{tabularx}
\usepackage{booktabs}
\usepackage{xcolor}
\usepackage{pifont}
\usepackage{amsthm}
\usepackage{extarrows}
\usepackage{hyperref}
\usepackage{balance}
\usepackage{subcaption}
\usepackage{caption}
\usepackage[ruled,vlined]{algorithm2e}

\usepackage{array}

\SetKwInput{KwInput}{Input}                
\SetKwInput{KwOutput}{Output}              

\def\x{{\mathbf x}}
\def\w{{\bm{\theta}}}
\def\ws{{\bm{\theta}_{\mathcal{S}}}}

\def\s{{\mathbf s}}

\def\L{{\cal L}}
\def\D{{\cal D}}
\def\R{{\mathbb R}}
\def\E{{\mathbb E}}

\def\R{\mathbb{R}}

\def\es{\widehat{\mathbf{s}}}

\def\ni{\mathbf{n}_{i}}
\def\n{\mathbf{n}}
\def\enone{\widehat{\mathbf{n}}_{1}}
\def\entwo{\widehat{\mathbf{n}}_{2}}
\def\eni{\widehat{\mathbf{n}}_{i}}
\def\en{\widehat{\mathbf{n}}}

\def\s{\mathbf{s}}

\def\mix{\mathbf{m}}

\def\tmix{\widetilde{\mathbf{m}}}

\def\tes1{\widetilde{\mathbf{s}}_{1}}
\def\tes2{\widetilde{\mathbf{s}}_{2}}

\def\tes{\widetilde{\mathbf{s}}}

\def\ten1{\widetilde{\mathbf{n}}_{1}}
\def\ten2{\widetilde{\mathbf{n}}_{2}}

\def\ten{\widetilde{\mathbf{n}}}

\makeatletter
\def\hlinewd#1{%
  \noalign{\ifnum0=`}\fi\hrule \@height #1 \futurelet
   \reserved@a\@xhline}
\makeatother

\title{Continual self-training with bootstrapped remixing \\ for speech enhancement}
\name{Efthymios Tzinis$^{1,*}$, \thanks{$^*$Work done during an internship at Meta Reality Labs Research.}Yossi Adi$^2$, Vamsi K. Ithapu$^3$, Buye Xu$^3$, Anurag Kumar$^3$}

\address{$^1$University of Illinois at Urbana-Champaign, $^2$Meta AI Research, $^3$Meta Reality Labs Research\\
\texttt{etzinis2@illinois.edu,\{adiyoss, ithapu, xub, anuragkr\}@fb.com}}
\begin{document}
\ninept
\maketitle
\begin{abstract}
We propose \textit{RemixIT}, a simple and novel self-supervised training method for speech enhancement. The proposed method is based on a continuously self-training scheme that overcomes limitations from previous studies including assumptions for the in-domain noise distribution and having access to clean target signals. Specifically, a separation teacher model is pre-trained on an out-of-domain dataset and is used to infer estimated target signals for a batch of in-domain mixtures. Next, we bootstrap the mixing process by generating artificial mixtures using permuted estimated clean and noise signals. Finally, the student model is trained using the permuted estimated sources as targets while we periodically update teacher's weights using the latest student model. Our experiments show that \textit{RemixIT} outperforms several previous state-of-the-art self-supervised methods under multiple speech enhancement tasks. Additionally, \textit{RemixIT} provides a seamless alternative for semi-supervised and unsupervised domain adaptation for speech enhancement tasks, while being general enough to be applied to any separation task and paired with any separation model.

\end{abstract}

\section{Introduction}
\label{sec:intro}
Neural networks have been found to be highly effective and widely applicable to a large number of audio and speech problems, including speech enhancement, where the goal is to improve the quality and intelligibility of degraded speech signals. In recent years, several neural architectures have reported state-of-the-art results for supervised \cite{pandey2021dense}, real-time \cite{hao2021fullsubnet} and semi-supervised \cite{Isik2020PoCONetSpeechEnhacement} speech enhancement tasks. Mostly driven by supervised learning, training such models requires large amounts of audio data which are expected to closely match the distribution of the test time input noisy recordings. While limited supervised data might often be available, supervised speech enhancement systems trained on them provides severely inferior performance due to the mismatch with the actual test distribution. 

To address these problems and to reduce reliance on purely supervised data, several speech enhancement and audio source separation studies have shifted their focus towards self-supervised methods \cite{xia2021incorporating}. 
 In \cite{sivaraman2021personalizedSE_SSL_dataPurification}, a model was trained to estimate the signal-to-noise ratio (SNR) of noisy mixture recordings and assign a confidence value on each noisy segment. Next, a separation model was trained on the noisy speech mixtures using a weighted reconstruction loss function to filter out the contribution of noisy ground-truth speech utterances.
Lately, mixture invariant training (MixIT) has been proposed in \cite{mixit} which enables unsupervised training of separation models by generating artificial mixtures of mixtures (MoMs) and letting the separation model estimate and reassign the sources back to the ground-truth mixtures. MixIT has been experimentally shown to provide a robust unsupervised solution for speech enhancement under a variety of setups \cite{mixit,fujimura2021noisyTargetTRaining,tzinis2021separate}. However, MixIT assumes access to in-domain noise samples and slightly alters the input SNR distribution by training on artificial MoMs with more than one noise sources \cite{saito2021trainingSEsystemsWNoisyDatasets}.

On the other hand, teacher-student setups have shown significant improvements in several audio processing tasks \cite{aihara2019teacherStudentDeepClustering}. In \cite{zhang2021teacher}, a student model was trained at the outputs of a pre-trained MixIT model for solving the problem of the artificially created input SNR mismatch between the train and test mixture distributions. An energy threshold was used to reduce the number of sources appearing in the noisy mixtures. Moreover, a student model can be adapted to a given test set using regression over the pre-trained teacher's estimates \cite{zeroshot_testTimeAdaptationSE}. Closest to our work is the self-training framework, proposed in \cite{wang2021semiSupSingVoiceSepNoisySelfTraining}, for semi-supervised singing voice separation where the teacher was pre-trained on an out-of-domain (OOD) supervised data and used for predicting estimated sources on the larger in-domain noisy dataset. The new self-labeled version of the dataset was filtered for low-quality separated sources and stored for offline training of a new student model using artificially generated mixtures from the estimated self-labeled estimated sources. Although all the aforementioned works assumed a frozen teacher, other works in automatic speech recognition have shown significant benefits when updating the teacher model using a moving mean teacher \cite{higuchi2021momentumPseudoLabelingforASR}.


In this paper, we propose a self-training method capable of performing self-supervised learning using large in-domain noisy datasets, while requiring only an OOD pre-trained teacher model (e.g. MixIT on an OOD dataset). In contrast to self-training methods in the literature which use ad-hoc filtering procedures to enhance the quality of the teacher model estimates, our method trains the student model by performing online remixing of the teacher's estimated sources. Moreover, instead of freezing the teacher model, \textit{RemixIT} treats the self-training process as a lifelong learning procedure by using sequential and moving averaging update schemes which enables faster convergence. Our experiments showcase the general applicability of our method towards \emph{self-supervised speech enhancement, semi-supervised OOD generalization}, and \emph{zero-shot domain adaptation}. We also provide an intuitive explanation of why our bootstrapped remixing process works under minimal assumptions.

\section{RemixIT Method}
\label{sec:remix_method}
We present \textit{RemixIT} for the general case of speech enhancement where the goal is to reconstruct the clean speech from a noisy speech signal. Formally, we train a separation model $f_{\mathcal{S}}$ which outputs $M$ sources for each noisy recording from an input batch $\x \in \R^{B \times T}$ containing $B$ waveforms, each with $T$ samples in the time-domain:
\begin{equation}
\label{eq:basic_estimation}
    \begin{gathered}
    \es, \en = f_{\mathcal{S}}(\x; \ws),\enskip
    \x = \s + {\textstyle \sum}_{i=1}^{M-1} \ni = \es + {\textstyle \sum}_{i=1}^{M-1} \eni , 
    \end{gathered}
\end{equation}
where $\es, \s \in \R^{B \times T}$, $\en, \n \in  \R^{(M-1) \times B \times T}$, $\ws$ are: the estimated speech signals, the clean speech targets, the estimated noise signals, the noise targets and the parameters of the model, respectively. In this work, we force the estimated sources $\es, \en$ to add up to the initial input mixtures $\x$ using a mixture consistency layer \cite{wisdom2019differentiableMixtureConsistency}.

\begin{algorithm}[t!]
\SetAlgoLined
    $\w^{(0)}_{\mathcal{T}} \gets \textsc{PretrainTeacher}(f_{\mathcal{T}}, \D')$ \\
    $\w_{\mathcal{S}} \gets \textsc{InitializeStudent}(f_{\mathcal{S}})$ \\
 \For{$k = 0$; $k{+}{+}$; while $k <= K$}{
 \For{\textsc{SampleBatch} $\mix \in \D_m, \enskip \mix \in \R^{B \times T}$}{ 
    $ \tes, \ten \gets f_{\mathcal{T}}(\mix; \w^{(k)}_{\mathcal{T}}) $ \tcp{Noisy estimates}
    $\tmix = \tes + \bm{\Pi} \ten$ \tcp{Bootstrapped remixing}
    $\es, \en  \gets f_{\mathcal{S}}(\tmix; \w^{(k)}_{\mathcal{S}}) $ \tcp{Student estimates}
    $\mathcal{L}_{\operatorname{RemixIT}} = \sum_{b=1}^B \left[ \L(\es_b, \tes_b) + \L(\en_b, \left[\bm{\Pi} \ten \right]_b) \right]$ \\
    $\w_{\mathcal{S}} \gets \textsc{UpdateStudent}(\w_{\mathcal{S}}, \nabla_{\w_{\mathcal{S}}} \mathcal{L}_{\operatorname{RemixIT}})$
 } 
 $\w^{(k+1)}_{\mathcal{T}} \gets \textsc{UpdateTeacher}(\w^{(k)}_{\mathcal{T}}, \w_{\mathcal{S}})$
}
 \caption{\textsc{RemixIT} for the noisy dataset $\D_m$.}
 \label{alg:remixit}
\end{algorithm}
 



\subsection{Mixture invariant training}
\label{sec:remix_method:mixit}
MixIT \cite{mixit} has proven its effectiveness under various self-supervised speech enhancement settings \cite{fujimura2021noisyTargetTRaining,tzinis2021separate}. Specifically, MixIT assumes that the training dataset consists of two portions $(\D_m, \D_n)$, where $\D_m$ is the part of the dataset which carries mixtures of speech and one noise source while $\D_n$ contains isolated noise recordings. During training, a new batch of artificial mixtures of mixtures (MoMs) is generated, $\x = \s + \mathbf{n}_1 + \mathbf{n}_2$, by sampling a batch of noisy speech recordings $\mix \sim \D_m$ and a batch of clean noise samples $\n_2 \sim \D_n$, where $\mix=\s + \n_1$. The separation model always estimates $M=3$ sources ($\es, \enone, \entwo = f_{\mathcal{S}}(\x; \ws)$) and the following permutation invariant loss is minimized for the $b$-th input MoM in the batch:
\begin{equation}
\label{eq:mixit_loss}
    \begin{aligned}
    \mathcal{L}_{\operatorname{MixIT}}^{(b)} = 
    \underset{ \bm{\pi} }{\min} 
    [ \L(\es_b + \en_{\pi_1, b}, \mathbf{m}_b) + \L(\en_{\pi_2, b}, \n_{2, b})  ], \enskip \forall b,
    \end{aligned}
\end{equation}
where $\bm{\pi} \in \{(2,3), (3,2)\}$ is the set of permutations of the output noise slots. However, MixIT depends on in-domain isolated noise recordings which makes it impractical for real-world settings. In most of the cases, matching the real noise distribution with the available noise set $\D_n$ is a strenuous task. The data augmentation proposed in \cite{saito2021trainingSEsystemsWNoisyDatasets} shows some improvements when an extra noise source from an OOD noise distribution is injected to the input MoM. Nevertheless, the performance of that method depends on the level of distribution shift between the actual noise distribution and $\D_n$.

\subsection{Continual self-training with bootstrapped remixing}
\label{sec:remix_method:remixit}
\textit{RemixIT} does not assume access to in-domain information. Thus, we can only draw mixtures from the in-domain noisy dataset $\mix = \s + \n$ ($m \sim \D_m$) where the noisy speech recordings contain a single noise source each and thus, $\mix, \s, \n \in \R^{B \times T}$. \textit{RemixIT} leverages a student-teacher framework to bootstrap the remixing process by permuting the previous noisy estimates, remixing them and using them as targets for training. We summarize \textit{RemixIT} in Algorithm \ref{alg:remixit}.
\subsubsection{RemixIT's continual self-training framework}
\label{sec:remix_method:remixit:student_teacher}
We assume that we can pre-train in a supervised or a self-supervised way a teacher model $f_{\mathcal{T}}$ on an OOD dataset $\D'$ which meets the specifications of Equation \ref{eq:basic_estimation}. Now, the first step is to use the teacher model to estimate some new noisy targets for a given mixture batch $\mix = \s + \n \in \R^{B \times T}, \enskip \mix \sim \D_m$ as follows:
\begin{equation}
\label{eq:teacher_inference}
    \begin{gathered}
    \tes, \ten = f_{\mathcal{T}}(\mix; \w^{(k)}_{\mathcal{T}}),\enskip
    \mix = \s + \n = \tes + {\textstyle \sum}_{i=1}^{M-1} \ten_i, 
    \end{gathered}
\end{equation}
where $k$ denotes the optimization step. If the teacher network was obtained by supervised (unsupervised via MixIT) OOD pre-training, we would have $M=2$ ($M=3$) output slots. Next, we use these estimated sources to generate new noisy mixtures as shown below:
\begin{equation}
\label{eq:bootstrapped_mixtures}
    \begin{gathered}
    \tmix = \tes + \ten^{(\bm{\Pi})} \in \R^{B \times T}, \enskip \ten^{(\bm{\Pi})} = \bm{\Pi} \ten, \enskip \bm{\Pi} \sim \mathcal{P}_{B \times B},
    \end{gathered}
\end{equation} 
where $\bm{\Pi}$ is drawn uniformly from the set of all $B \times B$ permutation matrices. Now, we simply use the permuted target pairs to train the student model $f_{\mathcal{S}}$ on the bootstrapped mixtures $\tmix$ as follows:
\begin{equation}
\label{eq:student_train}
    \begin{gathered}
    \es, \en = f_{\mathcal{S}}(\tmix; \w^{(k)}_{\mathcal{S}}), \enskip \es, \en \in \R^{B \times T} \\
    \mathcal{L}_{\operatorname{RemixIT}}^{(b)} = \L(\es_b, \tes_b) + \L(\en_b, \ten^{(\bm{\Pi})}_b), \enskip  b \in \{1, \dots, B\}, 
    \end{gathered}
\end{equation}
where the proposed loss function resembles a regular supervised setup with the specified signal-level loss function $\L$. Our method does not artificially alter the input SNR distributions similar to MixIT-like \cite{mixit,fujimura2021noisyTargetTRaining,saito2021trainingSEsystemsWNoisyDatasets} training recipes. Instead, the student model is trained on mixtures with the same number of sources for the bootstrapped mixtures where the teacher model had performed adequately. Unlike previous teacher-student methods which use the same teacher-estimated source-pairs as the targets for the student network \cite{zhang2021teacher,zeroshot_testTimeAdaptationSE}, the proposed bootstrapped mixtures increase the input mixture diversity and allow faster model training. This is especially useful in settings with a large distribution shift between teacher's and student's training data. Moreover, in contrast to the self-training approach in \cite{wang2021semiSupSingVoiceSepNoisySelfTraining}, where the teacher model is frozen and the inference on the in-domain dataset $\D_m$ is performed offline and the new self-labeled dataset is stored, \textit{RemixIT} employs a lifelong learning process. Our method is general enough that could be paired with any online co-training method which continuously updates the teacher's weights in addition to the main student training.


\subsubsection{Error analysis under the Euclidean norm}
\label{sec:remix_method:remixit:error_analysis}
We can express the errors produced by the student $\widehat{\mathbf{R}}_{\cal S}$ and the teacher $\widetilde{\mathbf{R}}_{\cal T}$ w.r.t. the initial clean targets as random variables:
\begin{equation}
\label{eq:student_teacher_errors}
    \begin{gathered}
    \widehat{\mathbf{R}}_{\cal S} = \widehat{\mathbf{S}} - \mathbf{S}, 
    \enskip \widetilde{\mathbf{R}}_{\cal T} = \widetilde{\mathbf{S}} - \mathbf{S},  \enskip (\mathbf{S},  \mathbf{N}) \sim \D  \\ 
    \widehat{\mathbf{R}}_{\cal S} \sim P(\widehat{\mathbf{R}}_{\cal S} | \widetilde{\mathbf{S}}, \widetilde{\mathbf{N}}, \bm{\Pi}), \enskip 
    \widetilde{\mathbf{R}}_{\cal T} \sim P(\widetilde{\mathbf{R}}_{\cal T} | \mathbf{S}, \mathbf{N}).
    \end{gathered}
\end{equation}
Now, we focus on the part of the objective function which is minimized at every student optimization step w.r.t. the speech component. Assuming unit-norm vector signals and using a signal-level loss $\L$ that minimizes the squared error between the estimated and target signals, the student's \textit{RemixIT} loss function is equivalent to:
\begin{equation}
\label{eq:equiv_student_train}
    \begin{aligned}
    \mathcal{L}_{\operatorname{RemixIT}} & \propto 
    \E \left[ || \widehat{\mathbf{S}} - \widetilde{\mathbf{S}} ||^2_2 \right] = 
    \E \left[ || ( \widehat{\mathbf{S}} - \mathbf{S} ) - ( \widetilde{\mathbf{S}} - \mathbf{S} ) ||^2_2 \right] \\
    & \propto 
    \underbrace{\E \left[ || \widehat{\mathbf{R}}_{\cal S} ||^2_2 \right]}_{\text{Supervised Loss}}
    + \underbrace{\E \left[ || \widetilde{\mathbf{R}}_{\cal T} ||^2_2 \right]}_{\text{Constant w.r.t. }\w_{\cal S}}
    - 2 \underbrace{\E \left[ \langle \widehat{\mathbf{R}}_{\cal S}, \widetilde{\mathbf{R}}_{\cal T} \rangle \right]}_{\text{Errors correlation}}.
    \end{aligned}
\end{equation}
Ideally, this loss could lead to the same optimization objective with a supervised setup if the last inner-product term was zero. $\langle \widehat{\mathbf{R}}_{\cal S}, \widetilde{\mathbf{R}}_{\cal T} \rangle = 0$ could be achieved if the teacher produced outputs indistinguishable from the clean target signals or the conditional error distributions in Equation \ref{eq:student_teacher_errors} were independent. Intuitively, as we continually update the teacher model and refine its estimates, we minimize the norm of the teacher error. Additionally, the bootstrapped remixing process forces the errors to be more uncorrelated since the student tries to reconstruct the same clean speech signals $\s$, similar to its teacher, but under a different mixture distribution. Formally, the student tries to reconstruct $\s$ when observing the bootstrapped mixtures $\widetilde{\mix} = \tes + \widetilde{\n}^{(\bm{\Pi})}$ while the teacher tries to reconstruct $\s$ from the initial input mixtures $\mix=\s+\n$. 


\section{Experimental Framework}
\label{sec:exp_fram}

\subsection{Datasets}
\label{sec:exp_fram:datasets}

\textbf{DNS-Challenge (DNS)}: We use the DNSChallenge 2020 benchmark dataset \cite{dnschallenge_2020} which covers a wide variety of noisy speech conditions. This dataset consists of $64{,}649$ and $150$ pairs of clean speech and noise recordings for training and testing, respectively. DNS is mainly used for showing the effectiveness of \textit{RemixIT} at leveraging vast amounts of noisy mixture recordings.

\noindent \textbf{LibriFSD50K (LFSD)}: This dataset consists of a diverse set of speakers drawn from the LibriSpeech \cite{librispeech} corpus and a wide variety of background noises from FSD50K \cite{fsd50k}. Specifically, $45{,}602$ and $3{,}081$ for training and validation, correspondingly. We follow the same data generation procedure as indicated in \cite{tzinis2021separate}. In this study, this dataset is mainly used for the OOD unsupervised or semi-supervised pre-training of speech enhancement models.

\noindent \textbf{WHAM!}: We follow the same procedure as in \cite{tzinis2021separate} in order to generate noisy mixtures using speakers and noise sources from the WHAM! \cite{WHAM} dataset. We use this dataset as a medium-sized dataset with $20{,}000$ training noisy-speech pairs and $3{,}000$ test mixtures. 

\noindent \textbf{VCTK}: We use the test partition of the VCTK dataset proposed in \cite{giri2019attention_VCTKDEMAND_testSet} which includes $586$ randomly generated noisy speech samples by mixing recordings from the VCTK speech corpus \cite{vctk} and the DEMAND \cite{demand_Dataset} noisy data collection.

\subsection{Separation model}
\label{sec:exp_fram:model}
We want to underline that the proposed method can be applied alongside any separation model. In this work, we chose the Sudo rm -rf~\cite{tzinis2020sudo} architecture since it achieves a good trade-off between speech enhancement quality and time-memory computational requirements. Specifically, we use the Sudo rm -rf variation with group communication \cite{luo2021ultra} and the default parameters in \cite{tzinis2021compute}, which has shown promising results in speech enhancement tasks \cite{tzinis2021separate}. All models have $U=8$ U-ConvBlocks except for the experiments where we increase the depth of the new student networks to $16$ and $32$. We fix the number of output slots to $M=3$ for MixIT models or $M=2$ otherwise.

\subsection{\textit{RemixIT} configurations}
\label{sec:exp_fram:remixit_coinfigs}
For the unsupervised \textit{RemixIT}, we assume that the initial teacher model was pre-trained using MixIT at a specified OOD dataset. For semi-supervised \textit{RemixIT}, we pre-train the teacher using conventional supervised training. We also experiment with various online teacher updating protocols such as:
\begin{equation}
\label{eq:update_protocols}
    \begin{gathered}
    \mathring{\w}^{(K)}_{\mathcal{T}} \coloneqq \w^{(K)}_{\mathcal{S}}, \enskip \bar{\w}^{(k+1)}_{\mathcal{T}} \coloneqq \gamma \w^{(k)}_{\mathcal{S}} + (1 - \gamma) \bar{\w}^{(k)}_{\mathcal{T}},
    \end{gathered}
\end{equation}
where $k$ denotes the training epoch index. For the sequentially updated teacher, we replace the old teacher with the latest student every $K=20$ epochs. For the zero-shot domain adaptation experiments, we first set the student to be the same as the teacher $\w^{(0)}_{\mathcal{S}} \coloneqq \bar{\w}^{(0)}_{\mathcal{T}}$ and then use the moving average teacher update with $\gamma=0.01$.



\subsection{Training and evaluation details}
\label{sec:exp_fram:train_eval_details}

Although we could use any valid signal-level loss function (see Equations \ref{eq:mixit_loss}, \ref{eq:student_train}), we choose the negative scale-invariant signal to distortion ratio (SI-SDR) \cite{sisdr}: 
\begin{equation}
\label{eq:SISDR}
    \begin{gathered}
    \mathcal{L}(\widehat{y}, y) = - \text{SI-SDR}(\widehat{y}, y) = - 10 \log_{10} \left( \nicefrac{\| \alpha y\|^2}{\| \alpha y - \widehat{y}\|^2} \right),
    \end{gathered}
\end{equation}
where $\alpha =  \widehat{y}^\top  y /\|y\|^2$ makes the loss invariant to the scale of the estimated source $\widehat{y}$ and the target signal $y$. We train all models using the Adam optimizer \cite{adam} with a batch size of $B=2$ and an initial learning rate of $10^{-3}$ which is divided by $2$ every $6$ epochs.

We evaluate the robustness of our speech enhancement models on the each test set after $100$ epochs for the supervised trained and pre-trained teachers as well as $60$ epochs for all the other configurations. Specifically, we report the SI-SDR \cite{sisdr}, the Short-Time Objective Intelligibility (STOI) \cite{stoi} and the Perceptual Evaluation of Speech Quality (PESQ) \cite{pesq} for $16$kHz target signals.

\section{Results}
\label{sec:results}

\begin{table*}[ht!]
    \centering
    \begin{tabular}{ll|c|cc|cc|cc|ccc}    
    \hlinewd{1pt}
    \multicolumn{2}{l|}{\multirow{3}{*}{Training Method and Model Details}}  & $\#$Model  & \multicolumn{6}{c|}{Available Training Data (\%)} & \multicolumn{3}{c}{Evaluation Metrics}  \\
    & &  Params & \multicolumn{2}{c|}{Clean Speech $\D_s$} & \multicolumn{2}{c|}{Clean Noise $\D_n$} & \multicolumn{2}{c|}{Mixture $\D_m$} & SISDR & \multirow{2}{*}{PESQ}  & \multirow{2}{*}{STOI}  \\
    & &  ($10^6$)  & DNS & LFSD & DNS & LFSD & DNS & LFSD & (dB) & &  \\
    \hlinewd{1pt}
    \multicolumn{2}{c|}{Input Noisy Mixture} & - & & & & & & & 9.2 & 1.58 & 0.915 \\
    \hlinewd{1pt}
    \multirow{3}{2.12cm}{Unsupervised MixIT with Student ($U=8$)} & In-domain & $0.79$ & & & 20\% & & 80\% & & 14.4 & 2.13 & 0.933 \\
    & OOD noise & $0.79$ & & &  & 20\% & 100\% & & 14.3 & 2.02 & 0.933 \\
    & Extra noise \cite{saito2021trainingSEsystemsWNoisyDatasets} & $0.79$ & & &  & 50\% & 100\% & & 14.5 & 2.03 & 0.930 \\
    \hlinewd{1pt}
    \multirow{2}{2.12cm}{Unsupervised RemixIT} & Teacher ($U=8$) & $0.79$ & & & & 20\% & & 80\% & 14.8 & 2.15 & 0.940 \\
    & Student ($U=32$) & $0.97$ & & & & & 100\% & & 16.0 & 2.34 & 0.952 \\
    \hlinewd{1pt}
    \multirow{2}{2.12cm}{Semi-supervised RemixIT} & Teacher ($U=8$) & $0.56$ & & 100\% & & 100\% & & & 17.6 & 2.61 & 0.958 \\
    & Student ($U=32$) & $0.73$ & &  & & & 100\% & & 18.0 & 2.60 & 0.959 \\
    \hlinewd{1pt}
    Supervised & Student ($U=8$) & $0.56$ & 100\% & & 100\% & & & & \textbf{18.6} & 2.69 & \textbf{0.962} \\
    In-domain & FullSubNet \cite{hao2021fullsubnet} & $5.6$ & 100\% & & 100\% & & & & 17.3 & \textbf{2.78} & 0.961 \\
    \bottomrule
    \end{tabular}
    \caption{Speech enhancement performance on the DNS test set using the proposed \textit{RemixIT} methods, unsupervised MixIT approaches \cite{mixit,saito2021trainingSEsystemsWNoisyDatasets} and supervised in-domain training with the same Sudo rm -rf model ($U=8$) and the state-of-the-art \textit{FullSubNet} model in the literature \cite{hao2021fullsubnet}.}
    \label{tab:final_table}
\end{table*}

\subsection{Continuous refinement of teacher's estimates}
\label{sec:results:semi_sup}

\begin{figure}[htb!]
    \centering
      \includegraphics[width=\linewidth]{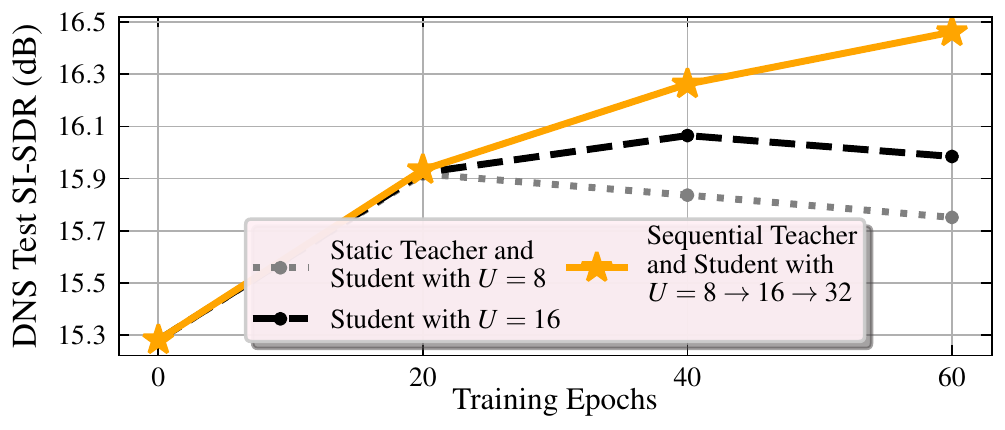}
      \caption{Speech enhancement performance on the DNS test set when using \textit{RemixIT} with different teacher update protocols. All approaches use the same teacher architecture with $U=8$ which was pre-trained in a supervised way using the training split of WHAM!. Notice that by sequentially updating the teacher (orange solid line) every $20$ epochs and replacing it with the previous student, we are able to obtain significant improvements over the methods which use the same static teacher (gray and black dashed lines). In this semi-supervised \textit{RemixIT} setup, considering the large mismatch between WHAM! and DNS datasets, the student model significantly outperforms the initial OOD pre-trained teacher by showing an improvement of more than $1$ dB in terms of SI-SDR performance.}
      \label{fig:static_vs_sequential}
\end{figure}

Our experimental results validate our hypothesis that a speech enhancement model could be trained faster and more effectively under a lifelong process where the teacher would be continuously updated in parallel to the student. The speech enhancement performance obtained by the sequentially updated and the frozen teacher protocols are shown in Figure \ref{fig:static_vs_sequential}. We notice that all protocols perform similarly until the $20$th epoch where the teacher is still static for all strategies. However, after the $20$th epoch, the teacher model is replaced with the latest student while the depth of the next student is increased $8 \rightarrow 16$. As a result, the student trained with the sequential method scales better for the same number of epochs (e.g. $40$th epoch) compared to the same size student ($U=16$) with a frozen teacher. We have experimentally seen that the sequentially updated teacher scales better than other protocols and this is the default strategy which we use across all other experiments except for the zero-shot adaptation where we also show that the running mean teacher updating scheme is also an effective option.

\subsection{Self-supervised and Self-supervised speech enhancement}
\label{sec:results:self_sup}
The speech enhancement results of the proposed method alongside supervised and unsupervised baselines are summarized in Table \ref{tab:final_table}. The percentage of the available data denotes the portion of each disjoint splits from the DNS or the LFSD paired data collections. For instance, the unsupervised \textit{RemixIT} teacher pre-training requires unsupervised MixIT using $80\%$ of the LFSD data pairs to simulate the noisy recordings $\D'_m$ and the other $20\%$ for the clean OOD noise recordings $\D'_n$, whilst the regular student training leverages the whole noisy DNS dataset.

Notice that unsupervised and semi-supervised \textit{RemixIT} does not depend on clean in-domain noise samples. Despite that, the unsupervised student model significantly outperforms all MixIT-like approaches including the in-domain training and the recently proposed extra noise augmentation where MoMs contain $3$ noise sources \cite{saito2021trainingSEsystemsWNoisyDatasets} ($14.5$dB $\rightarrow 16.0$dB in terms of SI-SDR). Moreover, the largest unsupervised student ($U=32$) outperforms its OOD MixIT unsupervised teacher by a large margin across all evaluation metrics which shows the efficacy of \textit{RemixIT} for self-supervised settings. The proposed method also yields noticeable gains for the semi-supervised case where the student model performs comparably with in-domain supervised training using the default Sudo rm -rf model with $U=8$ and a recent state-of-the-art model. We want to underline that our method could be used with more complex models as teachers, rather than the efficient Sudo rm -rf architecture, and provide even higher quality speech enhancement performance.

\subsection{Zero-shot domain adaptation}
\label{sec:results:zero_shot}
\begin{figure}[ht]
    \centering
      \includegraphics[width=\linewidth]{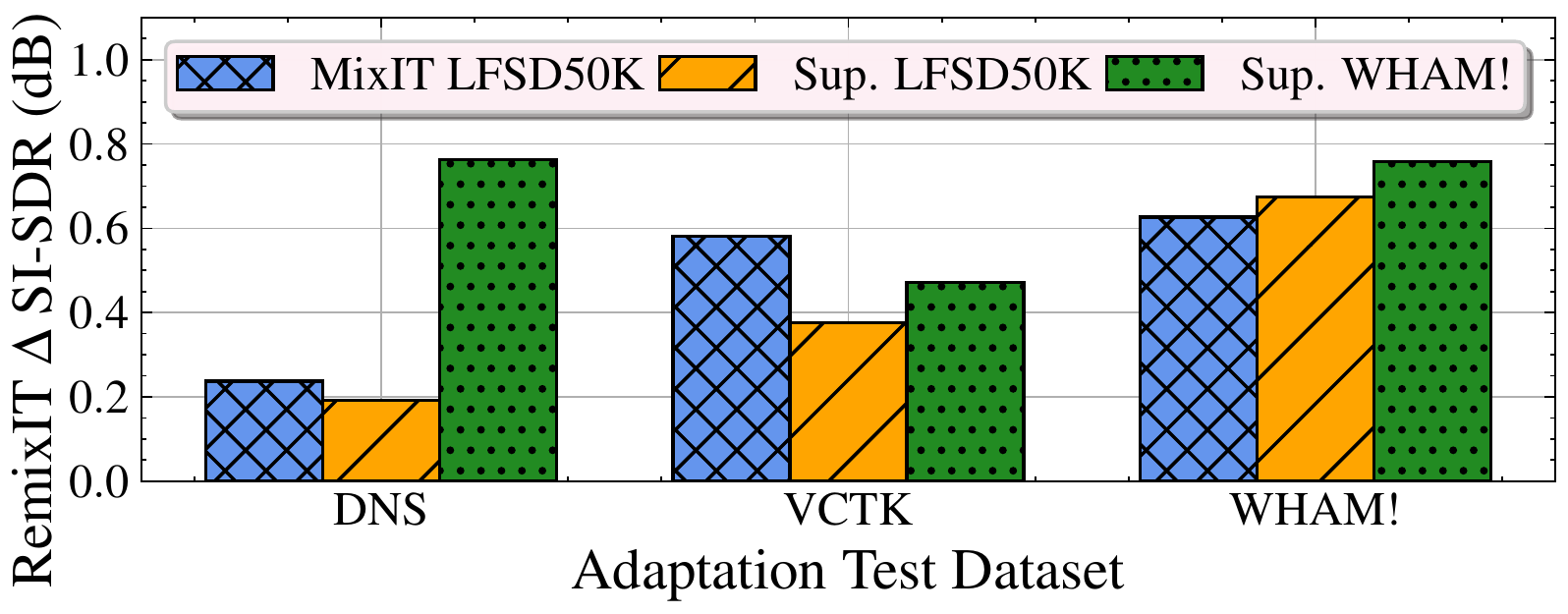}
      \caption{Absolute gain in terms of SI-SDR when using \textit{RemixIT} with an OOD pre-trained model and a given test set to adapt (e.g. DNS, LFSD and WHAM!, from left to right). We show our results on three different pre-trained Sudo rm -rf models with $U=8$ and following MixIT unsupervised pre-training on LFSD (blue/leftmost), supervised training on LFSD (yellow) and supervised pre-training on WHAM! (green/rightmost).}
      \label{fig:zeroshot}
\end{figure}

We show that \textit{RemixIT} can also be used with low-resource datasets, where training data are limited but one has access to a test dataset for adapting a pre-trained model. In Figure \ref{fig:zeroshot}, the performance improvement for the zero-shot speech enhancement task is depicted with a variety of supervised and unsupervised pre-trained networks on larger OOD datasets. Notably, \textit{RemixIT} yields improvements of up to $0.8$dB in terms of SI-SDR compared to the uncalibrated pre-trained models while using a limited amount of in-domain mixtures. The performance of our model is correlated with the amount of available noisy mixtures and this is the reason we see the largest (smallest) gains for the WHAM! (DNS) test partition which has $3{,}000$ (only $150$) mixtures. Moreover, we also notice a large improvement in cases where there is a large distribution shift between the training data and the mixtures in the adaptation set (e.g. supervised training on WHAM! and adapting on the relatively small DNS test set).

\section{Conclusion}
\label{sec:conclusion}
We have proposed a novel continual self-training method for denoising and have experimentally showed its benefits on several realistic speech enhancement tasks. Our method depends only on the existence of in-domain noisy data and a pre-trained model using purely out-of-domain data which might not necessarily capture the in-domain distribution. The coupling of the bootstrap remixing process with the continuously bi-directional teacher-student self-training framework leads to significant improvements for zero-shot and self-supervised speech enhancement as well as semi-supervised domain adaptation. In the future, we aim to apply our method to other domains and denoising tasks as well as provide stronger theoretical guarantees for the convergence of our algorithm.




\bibliographystyle{IEEEbib}
\bibliography{refs}

\end{document}